\documentclass[a4paper,11pt]{article}
\usepackage{fontenc}
\usepackage{makeidx}
\usepackage{amssymb}

\title{Conserved Charges in the Principal Chiral Model on a Supergroup}
\author{B.H. Miller,\\ \textit{Department of Mathematics, University of York},\\
\textit{Heslington Lane, York YO10 5DD, UK}\\Email:
bhm101@york.ac.uk}
\date{}

 \usepackage{amsfonts}

\begin{document}
\maketitle

\begin{abstract}
The classical principal chiral model in 1+1 dimensions with target
space a compact Lie supergroup is investigated. It is shown how to
construct a local conserved charge given an invariant tensor of the
Lie superalgebra. We calculate the super-Poisson brackets of these
currents and argue that they are finitely generated. We show how to
derive an infinite number of local charges in involution. We
demonstrate that these charges Poisson commute with the non-local
charges of the model.
\end{abstract}
\section{Introduction}

Finding exact methods of solving string theory on AdS space, as
motivated by the AdS/CFT correspondence, remains a difficult
problem. There are two schemes for describing such a superstring.
The NSR description gives a free action in a flat background but the
existence of the RR vertex operators introduces seemingly
insurmountable difficulties. Alternatively there is the GS formalism
in which the supersymmetry exists on the target manifold which is
then described as a Lie supergroup or supercoset space. \vspace{1em}

It has proved possible to describe superstring theory on
$AdS_n\times S^n$ as a coset space $G/H$ (for example, $AdS_2 \times
S^2$ is the bosonic subalgebra of $\frac{PSU(1,1|2)}{U(1)\times
U(1)}$ \cite{ads2}, and $AdS_5 \times S^5$ is the same for
$\frac{PSU(2,2|4)}{SO(4,1)\times SO(5)}$ \cite{ads5, transition
matrices, metsaev}.) Although $G$ is Ricci flat in both these
instances, the coset space is not. However, $H$ is the invariant
locus of a $\mathbb{Z}_4$ automorphism, and this permits the
introduction of a WZ-term to provide a quantum conformal theory.
(Indeed \cite{Zm grading, Young Kagan} have demonstrated quantum
conformal invariance given a $\mathbb{Z}_n$ automorphism.) Similarly
superstring theory on $AdS_3 \times S^3$ is related to a sigma model
on $PSU(1,1|2)$ \cite{CFT of AdS with RR flux}. More general work
has looked at sigma models on the supergroups $PSL(n|n)$ \cite{psl
cft}.\vspace{1em}

All of these models contain an infinite number of local and
non-local conserved charges. Their existence constrains the S-matrix
and permits its exact computation. These charges have been studied
in both bosonic and worldsheet supersymmetric principal chiral
models (PCMs) \cite{local conserved charges}, and for sigma models
on symmetric spaces \cite{evans mountain, quantum symmetric
models}.\vspace{1em}

The PCM on any Lie supergroup is classically conformal. In the
quantum model, the one loop beta function is proportional to the
dual Coxeter number $h^{\vee}$ \cite{psl cft, saleur kaufmann}, and
there are some superalgebras for which this vanishes (namely
$psu(n|n)$ and $osp(2n+2|2n)$). The PCM on these supergroups will
therefore be quantum conformal \cite{psl cft, polchinkski
supercoset}. However we do not expect conformal invariance to
survive in the quantum model for general supergroups when the model
becomes massive. This paper is a modest attempt at understanding the
algebra of these models' local charges, with the hope that some
insight could be given to the perhaps more physical
models.\vspace{1em}

Ultimately we find that the classical PCM on a supergroup has an
infinite number of charges in involution. These charges are formed
from integrals of local conserved currents, each associated to an
invariant of the Lie superalgebra. Only a finite number of these are
independent, and from these invariants it is possible to construct
all conserved local currents.\vspace{1em}

For the $su(m|n)$ models we will construct a set of currents which
give rise to commuting charges. The construction will fail for
$m=n$, so we restrict ourselves to the massive cases. The situation
is analogous for $osp(m|2n)$, except now a family of currents
exists, depending on a free parameter $\alpha$. For $osp(2m|2n)$ an
additional invariant exists, the superpfaffian \cite{superpfaffian},
and requiring that the superpfaffian charge commutes with the other
local charges fixes the value of $\alpha$, but only when the model
is massive.\vspace{1em}

These charges are conserved classically,  but anomalies might arise
in the quantum model. However we anticipate that integrability
survives because we can use Goldschmidt and Witten's method of
anomaly counting \cite{Goldschmidt Witten} to show that higher spin
conservation laws still exist.\vspace{1em}

In addition to the local charges, there are non-local charges which
form a Super-Yangian structure \cite{superyangian on ads5, zhang,
yangian symmetry in d=4 SYM}. We shall show that these non-local
charges are in involution with the local charges.\vspace{1em}


\subsection{Lie Superalgebras}

We begin with a basic introduction to superalgebras, and refer to
the literature \cite{sorba, Kac, scheunert} for a more thorough
review. Throughout this paper, letters $A,B,C,\ldots$ will denote
both bosonic and fermionic indices. $a,b,c,\ldots$ will be used for
bosonic indices, and $\alpha, \beta, \gamma, \ldots$ for fermionic
indices.\vspace{1em}

We begin with an associative Grassmann algebra
$\Lambda=\Lambda_{\bar{0}}\oplus\Lambda_{\bar{1}}$ with sufficiently
many anticommuting generators, where $\Lambda_{\bar{0}}$ (resp.
$\Lambda_{\bar{1}}$) consists of commuting (resp. anticommuting)
elements. We have the product rule $\Lambda_{\bar{i}}\cdot
\Lambda_{\bar{j}}\subset \Lambda_{\overline{i+j}}$. (Addition modulo
2 is left implicit here and throughout.)\vspace{1em}

Given a supermatrix $X=\left(\begin{array}{cc} A & B\\C &D
\end{array}\right)$ we define it to be even (odd) if
$A,D\in\Lambda_{\bar{0}}$ ($\Lambda_{\bar{1}}$) and
$B,C\in\Lambda_{\bar{1}}$ ($\Lambda_{\bar{0}}$). We write
$\textmd{deg}(X)=0$ if it is even and $\textmd{deg}(X)=1$ if odd. We
can then define the supertrace as
\begin{equation}
\textmd{Str}(X)=\textmd{Tr}(A)-(-1)^{\textmd{deg}(X)}\textmd{Tr}(D)
\end{equation}

the supertranspose as
\begin{equation}
X^{ST}=\left(\begin{array}{cc} A^T &
-(-1)^{\textmd{deg}(X)}B^T\\(-1)^{\textmd{deg}(X)}C^T &D^T
\end{array}\right)
\end{equation}

and if $X$ is even and invertible, the superdeterminant as
\begin{equation}
\textmd{sdet}(X)=\frac{\textmd{det}(A-BD^{-1}C)}{\textmd{det}(D)}=
\frac{\textmd{det}(A)}{\textmd{det}(D-CA^{-1}B)}.
\end{equation}

These satisfy the important properties
\begin{equation}
(XY)^{ST}=(-1)^{\textmd{deg}(X)\textmd{deg}(Y)}Y^{ST}X^{ST}
\end{equation}
\begin{equation}
\textmd{Str}(XY)=(-1)^{\textmd{deg}(X)\textmd{deg}(Y)}\textmd{Str}(YX)\;
, \; \textmd{Str}(X^{ST})=\textmd{Str}(X)
\end{equation}
\begin{equation}
\textmd{sdet}(XY)=\textmd{sdet}(X)\textmd{sdet}(Y)\; , \;
\textmd{sdet}(X^{ST})=\textmd{sdet}(X)
\end{equation}
\begin{equation}
\textmd{sdet}(\exp(X))=\exp(\textmd{Str}(X))
\end{equation}

We consider a compact connnected supergroup $G$ \cite{sorba}, either
$SU(m|n)$ or the compact subgroup of $OSp(m|2n)$ which is connected
to the identity, satisfying
\begin{eqnarray}
SU(m|n)\;:& \textmd{sdet} X = 1 & XX^{\dag}=1\\
OSp(m|2n)\;:& \textmd{sdet}(X)=1& X^{ST}HX=H
\end{eqnarray}

where $H=\left(\begin{array}{cc}I & 0\\0 & J \end{array}\right) $
for an symmetric $I$ and symplectic $J$. There exists a basis for
which $I$ is the identity $\mathbb{I}_m$ and $J=
\left(\begin{array}{cc}0 & \mathbb{I}_{n}\\ -\mathbb{I}_{n} &
0\end{array}\right) $.\vspace{1em}

The relation between a supergroup and its superalgebra is similar to
the bosonic case. The supergroup $G$ associated to the superalgebra
$\mathbf{g}$ is the exponential mapping of the even subsuperalgebra
of the Grassmann envelope $\Lambda \otimes \mathbf{g}$.
\begin{equation}\label{exp map}
g=\exp(x_AT^A)=\exp(x_aT^a)\exp(x_{\alpha}T^{\alpha})
\end{equation}

$T^A$ generate the Lie superalgebra
$\mathbf{g}=\mathbf{g}_{\bar{0}}\oplus \mathbf{g}_{\bar{1}}$
corresponding to $G$ (where $T^a$ (respectively $T^{\alpha}$)
generate $\mathbf{g}_{\bar{0}}$ (respectively
$\mathbf{g}_{\bar{1}}$). These satisfy the supercommutation
relationship
\begin{equation}
[T^A,T^B]=T^AT^B-(-1)^{\eta_A \eta_B}T^BT^A=f^{AB}_{\phantom{AB}C}
T^C
\end{equation}

for structure constants $f^{AB}_{\phantom{AB}C}$. Note that the even
subspace $\mathbf{g}_{\bar{0}}$ is itself a Lie algebra, whereas
$\mathbf{g}_{\bar{1}}$ is a
$\mathbf{g}_{\bar{0}}$-module.\vspace{1em}

We denote by $\eta_A$ the grade of $T^A$ in the Lie superalgebra,
and note that the structure constants are antisupersymmetric in the
first two indices.
\begin{equation}
f^{AB}_{\phantom{AB}C}=-(-1)^{\eta_A \eta_B}f^{BA}_{\phantom{AB}C}
\end{equation}

Furthermore, using the invariance of the supertrace
\begin{equation}
\textmd{Str}(T^A[T^B,T^C])=\textmd{Str}([T^A,T^B]T^C)
\end{equation}

we find that the structure constants are also antisupersymmetric in
the first and last indices.
\begin{equation}
f^{AB}_{\phantom{AB}C}=-(-1)^{\eta_A \eta_C}f^{CB}_{\phantom{AB}A}
\end{equation}

The $x_A$ commute (resp. anticommute) whenever the $T^A$ are graded
even (resp. odd), and so we can define without ambiguity $\eta_A$ to
be the gradation of $x_A$ in the Grassmann algebra. The local
currents from the PCM are constructed from elements of this Lie
algebra.\vspace{1em}

We are interested in the Lie superalgebras $su(m|n)$, and
$osp(m|2n)$, satisfying
\begin{eqnarray}
su(m|n)&: &  \textmd{Str} (X)=0 \; , \;X=-X^{\dag}\\
osp(m|2n)& :& X^{ST}=-HXH^{-1}
\end{eqnarray}

For the Lie superalgebras $su(m|n)$ $(m\neq n)$ and $osp(m|2n)$ we
can define a non-degenerate invariant bilinear form
$G^{AB}=\textmd{Str}(T^A T^B)$. Note that this implies that
$G^{AB}=0$ unless $\eta_A=\eta_B$ (consistency) and that
$G^{AB}=(-1)^{\eta_A}G^{BA}$ (supersymmetry). We define the inverse
$G_{AB}$ via
\begin{equation}
G_{AB}G^{BC}=\delta_A^{\; C}
\end{equation}

Note in particular that this implies that
\begin{equation}
G^{AB}X_AY_B=X_AY^A=(-1)^{\eta_A}X^AY_A=G_{BA}X^AY^B.
\end{equation}

There is a one-dimensional ideal of $su(n|n)$, generated by
$i\mathbb{I}_{2n}$. It is therefore not semi-simple and has no
non-degenerate metric, and it is usual to consider the PCM on
$PSU(n|n)=SU(n|n)/\mathbb{I}_{2n}$. The Killing form  for this is
exactly zero (it is proportional to the dual Coxeter number) but an
invariant bilinear form does exist. We shall avoid this by later
imposing $m\neq n$.\vspace{1em}

We will make much use of the completeness condition in what is to
follow. For any $X =X_A T^A \in \textbf{g}$ we have
\begin{equation}\label{completeness}
X_A = \textmd{Str}(T_A X).
\end{equation}


\section{Local Charges of the PCM}

The PCM is defined by the lagrangian
\begin{equation}
\mathcal{L}=\frac{\kappa}{2}\textmd{Str}(\partial_{\mu}g^{-1}\partial^{\mu}g)
\end{equation}

where $g$ takes values in a supergroup $G$, either $SU(m|n)$ or
$OSp(m|2n)$. $\kappa$ is a dimensionless constant, its value
unimportant for the classical model, and $\mu$ indexes the 1+1D
spacetime. This lagrangian is invariant under a global chiral
symmetry $g(x,t) \rightarrow g_1 g(x,t) g_2^{-1}$, with associated
Noether conserved local currents
\begin{equation}\label{Noether currents}
j_{\mu}^L=\kappa \partial_{\mu}gg^{-1} \quad , \quad
j_{\mu}^R=-\kappa g^{-1} \partial_{\mu}g.
\end{equation}

The conservation of these currents are the equations of motion.
These currents belong to the even subspace of
$\mathbf{g}\otimes\Lambda$.\vspace{1em}

From these conserved currents it is possible to construct the higher
spin local and non-local conserved charges. Using either the left or
right currents will give rise to identical local charges. The
non-local charges constructed from them are not equal, but form two
copies of a super-Yangian structure. When there is no confusion, we
shall drop the $L/R$ indices.\vspace{1em}

We assume appropriate boundary conditions on $j_{\mu}(x)$
\begin{equation}
j_{\mu}(x)=0 \quad \textmd{as } x\rightarrow \pm\infty.
\end{equation}

The currents are conserved, and satisfy the Bianchi identity
\begin{equation}\label{conserve_bianchi}
\partial^{\mu}j_{\mu}=0 \quad , \quad
\partial_{\mu}j_{\nu}-\partial_{\nu}j_{\mu}-\frac{1}{\kappa}[j_{\mu},j_{\nu}]=0.
\end{equation}

Immediately we can form conserved charges out of the Noether
currents (\ref{Noether currents}) by $Q^{(0)}=\int dx\;j_{0}$. More
importantly though, conditions (\ref{conserve_bianchi}) allow us to
form a local conserved charge of spin $s$ through the use of a
$G$-invariant tensor of degree $s+1$. Although there are infinitely
many such invariant tensors, there are only finitely many
independent (primitive) tensors. The number is equal to the rank of
$\mathbf{g}$. All other invariants are constructed from these
primitive ones.\vspace{1em}

The intention is to find a maximal set of mutually commuting
conserved local charges $\{q_s\}$. The existence of these currents
displays the integrable nature of the PCM, because (equivalent to
the construction of non-local charges) the two conditions in
(\ref{conserve_bianchi}) allow a Lax pair to be
constructed.\vspace{1em}

It will prove useful to write conditions (\ref{conserve_bianchi}) in
terms of light-cone coordinates, $x^{\pm}=\frac{1}{2}(t\pm x)$
\begin{equation}\label{lc conserve and flat}
\partial_-j_+=-\partial_+j_-=-\frac{1}{2\kappa}[j_+,j_-].
\end{equation}


\subsection{The Energy-Momentum Tensor and Conformal Invariance}

The energy-momentum tensor is the variation of the lagrangian with
respect to the spacetime metric,
\begin{equation}\label{em_tensor}
T_{\mu \nu}=-\frac{1}{2\kappa}\left(
\textmd{Str}(j_{\mu}j_{\nu})-\frac{1}{2}\eta_{\mu\nu}\textmd{Str}(j_{\rho}j^{\rho})\right)
\end{equation}

This is traceless, symmetric and conserved. In light-cone
coordinates
\begin{equation}
T_{\pm\pm}=-\frac{1}{2\kappa}\textmd{Str}(j_{\pm}j_{\pm})\;,\quad
T_{+-}=T_{-+}=0.
\end{equation}

and
\begin{equation}
\partial_-T_{++}=\partial_+T_{--}=0
\end{equation}

Here $T_{+-}$ is the trace of the two-dimensional energy-momentum
and its vanishing implies that the model has classical conformal
invariance. The situation is more complicated in the presence of
quantum anomalies. The one loop beta function is proportional to
$h^{\vee}$, the dual Coxeter number of the Lie (super)algebra
\cite{saleur kaufmann}. For purely bosonic Lie algebras,
$h^{\vee}\neq 0$, and so a WZ term must be added to the lagrangian
for conformal invariance in the quantum model \cite{local conserved
charges}. However there are some Lie superalgebras (namely
$psl(n|n)$ and $osp(2n+2|2n)$) for which $h^{\vee}=0$, and these
models retain conformal invariance in the quantum model, at least to
one loop \cite{psl cft, polchinkski supercoset}.\vspace{1em}

We note here that we can form a series of higher-spin conservation
laws,
\begin{equation}
\partial_-(T_{++}^p)=\partial_+(T_{--}^p)=0.
\end{equation}

These give the classical conformal symmetry of the model, but are
not expected to be preserved for the quantum models for which the
dual Coxeter number is non-zero. We shall not be concerned with
these directly however, as we shall see that more general
higher-spin currents can be formed.


\subsection{Canonical Formalism}

Our aim here is to calculate the super-Poisson brackets (SPBs) for
the current components $j_{\mu A}$ \cite{superpoisson}. As in the
bosonic case \cite{local conserved charges, faddeev takhtajan}, it
is convenient to introduce the non-local operator
$\Delta_1=\partial_1-\frac{1}{\kappa}[j_1,\;]$ using which the
Bianchi identity is re-expressed as $j_0 = \Delta_1^{-1} (\partial_0
j_1)$. We can now write the action as a functional of $j_1(x)$ only
\begin{equation}
\mathcal{L}=\frac{1}{2\kappa}\textmd{Str}((\Delta^{-2}_1\partial_0j_1)
(\partial_0j_1)-j_1^2)
\end{equation}

(where we have imposed suitable boundary conditions such that
$\Delta_1^{-1}(A)B=-A\Delta_1^{-1}(B)$ up to a total
divergence).\vspace{1em}

Defining the conjugate momentum of $j_1$ to be $\pi=\pi^AT_A$, where
$\pi^A=\partial\mathcal{L}/\partial(\partial_0j_{1A})$ we find that
$j_0=-2\kappa\Delta_1 \pi$. Then, using
$\{j_{1A}(x),\pi^B(y)\}=\delta_A^B\delta(x-y)$, we find that
\begin{eqnarray}\label{spacetime SPBs}
\{ j_{0 A}(x),j_{0 B}(y) \}&=&(-1)^{\chi}f_{AB}^{\phantom{AB}C}j_{0
C}(x)\delta(x-y)\nonumber\\
&&\nonumber\\ \{ j_{0 A}(x),j_{1 B}(y) \} & = &
(-1)^{\chi}f_{AB}^{\phantom{AB}C}j_{1 C}(x)\delta(x-y)\nonumber\\
& & +\kappa G_{AB}\partial_x\delta(x-y)\\
&&\nonumber\\ \{ j_{1 A}(x),j_{1 B}(y) \}&=&0\nonumber
\end{eqnarray}

where $\chi = \eta_A . \eta_B + \eta_A + \eta_B$. In light-cone
coordinates these become
\begin{eqnarray}\label{lightcone SPBs}
\{j_{\pm A}(x), j_{\pm B}(y) \}&=&
(-1)^{\chi}f_{AB}^{\phantom{AB}C}\left( \frac{3}{2} j_{\pm
C}(x)-\frac{1}{2}j_{\mp C}(x)\right)\delta(x-y)\nonumber\\
&&\pm 2\kappa G_{AB}\delta'(x-y)\\
&&\nonumber\\
\{j_{+ A}(x), j_{- B}(y) \}& =&
\frac{1}{2}(-1)^{\chi}f_{AB}^{\phantom{AB}C}\left[ j_{+ C}(x)+j_{-
C}(y)\right]\delta(x-y)\nonumber
\end{eqnarray}

These are very similar in form to the Poisson brackets of the
bosonic model \cite{local conserved charges} but we must take
account of the non-trivial gradings.\vspace{1em}


\subsection{Higher-Spin Conserved Charges}

We can construct the Noether $G_L \times G_R$ conserved charges
\begin{equation}
Q^{0}_A=\int^{\infty}_{-\infty}\textit{dx} \;j_{0 A}(x)
\end{equation}

for both left and right currents $j_{\mu}^L$ and $j_{\mu}^R$. More
interestingly we can use the above SPBs to find an infinite set of
commuting higher-spin holomorphic local currents. We shall obtain
the same set of local currents if we use either the left or right
current. We first note that, similar to the bosonic model
\cite{local conserved charges}, to every invariant tensor
$d^{A_1\ldots A_p}$ (supersymmetric in adjacent indices) associated
with a Casimir element of degree $p$, we can associate a conserved
current of spin $p$. Denote such a Casimir element
\begin{equation}\label{d polynomial}
\mathcal{C}^p=d^{A_1,\ldots A_p}T_{A_1}\ldots T_{A_p}
\end{equation}

and supersymmetry means that
\begin{equation}\label{susy tensor}
d^{A_1 \ldots A_k A_{k+1}\ldots
A_p}=(-1)^{\eta_{A_k}\eta_{A_{k+1}}}d^{A_1 \ldots A_{k+1}
A_{k}\ldots A_p}
\end{equation}

We note that, for $su(m|n)$ and $osp(m|2n)$, $d^{A_1,\ldots A_p}=0$
unless $\sum \eta_{A_i}=0$. Indeed, Casimir elements are bosonic for
all basic superalgebras except $Q(n)$ \cite{sorba}. Invariance then
implies that
\begin{equation}\label{d tensor}
[\mathcal{C}^p,T^B]=0 \quad \Rightarrow \quad
\sum_{i=1}^p(-1)^{\eta_B(\eta_{A_i}+\ldots \eta_{A_p})}d^{A_1 \ldots
\hat{A_i}C\ldots A_p}f^{A_i B}_{\;\;\;\;\;\;C}=0
\end{equation}

We define the action of this tensor on an element of the Lie
superalgebra $X=X_AT^A$ by
\begin{equation}
d^{(p)}(X)=d^{A_1 \ldots A_p}X_{A_1}\ldots X_{A_p}
\end{equation}

and observe that it is $G$-invariant

\begin{equation}
d^{(p)}(gXg^{-1})=d^{(p)}(X)\; \textmd{for } g\in G.
\end{equation}

We can simplify matters by noting that every element of a Lie
(super)algebra is locally conjugate to some element of its Cartan
subalgebra (CSA), $\mathbf{h}\subset\mathbf{g}$ i.e. its maximal
abelian subgroup. For most basic Lie superalgebras, $\mathbf{h}$ is
the CSA of its bosonic subalgebra
$\mathbf{h}\subset\mathbf{g}_{\bar{0}}$. \vspace{1em}

Using the invariance property we then have
\begin{equation}
d^{(p)}(X)=d^{(p)}(H)=d^{a_1\ldots a_p}H_{a_1}\ldots H_{a_p}
\end{equation}

where $H=H_aT^a=gXg^{-1}\in \mathbf{h}$. We can thus consider the
invariant tensor to be restricted to the CSA, and so we are only
interested in the invariants of the underlying bosonic Lie
subalgebra.\footnote{The exception is the strange superalgebra
$Q(n)$, where $\mathbf{h}\cap\mathbf{g}_{\bar{1}}\neq \emptyset$
\cite{sorba}.}\vspace{1em}

To each superalgebra there are an infinite number of invariant
tensors, notably of the form $\textmd{Str}(T^{A_1}\ldots T^{A_m})$,
but there are only finite many independent (or primitive) tensors,
the number being equal to the rank of the superalgebra. All local
conserved charges will be generated by charges formed from these
primitive tensors. These primitive tensors were discussed in
\cite{invariant simple, invariant compact} for the bosonic algebras,
and for superalgebras in \cite{Casimir invariants and characteristic
identities, invariant polynomials}\vspace{1em}

An infinite set of higher spin conserved charges can be constructed
from tensors which satisfy (\ref{d tensor}). Using (\ref{lc conserve
and flat}) it is straightforward to verify that
\begin{equation}
\partial_-(d^{A_1\ldots A_p}j_{+A_p}\ldots j_{+A_1})=0
\end{equation}

where the change in ordering of the indices is non-trivial because
of the $\mathbb{Z}_2$-grading (\ref{susy tensor}). Then the local
conserved charges are

\begin{equation}
q_{\pm s} = \int \textit{dx}\; d^{A_1 \ldots A_{s+1}}j_{\pm
A_{s+1}}(x)\ldots j_{\pm A_1}(x).
\end{equation}

The charges are labelled by $s$, their spin. The Poisson bracket
with the (purely bosonic) boost generator $M$ is $\{M,q_{\pm
s}\}=\pm s q_{\pm s}$. We shall define the charges $q_{\pm s}$ with
$s>0$ as having positive/negative chirality. The additive nature of
these local charges implies that they have a trivial coproduct
\begin{equation}
\Delta(q_{\pm s})=q_{\pm s}\otimes \mathbb{I} + \mathbb{I}\otimes
q_{\pm s}
\end{equation}
Using the invariance condition (\ref{d tensor}), it is simple to
show that $\{q_{+r} , q_{-s} \}=0$ for any integers $r,s>0$.
Furthermore, for charges of equal chirality, only the non-ultra
local terms contribute.
\begin{eqnarray}\label{d-tensor commutation}
\{q_{\pm r} , q_{\pm s} \}&=&\pm 2(r+1)(s+1)\kappa\int \textit{dx}
\; (-1)^{\eta_B(\eta_{B_1}+
\ldots +\eta_{B_s})}d^{A_1 \ldots A_r A}d^{B_1 \ldots B_s B}G_{AB}\nonumber\\
&&\times j_{\pm A_1}\ldots j_{\pm A_r} \partial_{x}(j_{\pm
B_1}\ldots j_{\pm B_s})
\end{eqnarray}

We are interested in the currents formed from a tensor with
components $d^{A_1 \ldots A_r}=\textmd{sStr}(T^{A_1}\ldots
T^{A_r})$, which satisfies (\ref{d tensor}). sStr denotes the
normalised supersymmetric supertrace
\begin{equation}
\textmd{sStr}(T^{A_1} T^{A_2} \ldots
T^{A_r})=\frac{1}{r!}\sum_{\sigma \in S_r}
\epsilon_{\sigma}\textmd{Str}(T^{A_{\sigma(1)}} T^{A_{\sigma(2)}}
\ldots T^{A_{\sigma(r)}})
\end{equation}

where $S_r$ is the symmetric group of degree $r$, and
$\epsilon_{\sigma}=-1$ if $\sigma$ involves an odd number of
permutations of fermionic indices, and equals 1 otherwise. This
tensor is exactly zero unless $\eta_{A_1}+\ldots + \eta_{A_r}=0$ mod
2. This gives rise to holomorphic currents $\mathcal{J}_{\pm
r}=\textmd{Str}(j_{\pm}^r)$, and associated conserved charges
\begin{equation}\label{supertrace higher charge}
q_{\pm (r-1)}=\int \textit{dx} \; \textmd{Str}(j_{\pm}^r),
\end{equation}

and we find that that equation (\ref{d-tensor commutation})
simplifies to

\begin{equation}\label{supertrace charge commutation}
\{ q_{\pm r},q_{\pm s}\}=\pm 2(r+1)(s+1)\kappa\int \textit{dx} \;
\textmd{Str} (j_{\pm}^r T_A)\partial_x \textmd{Str}(j_\pm^s T^A).
\end{equation}

We now want to find the currents $\mathcal{J}_r(x)$ (or indeed,
algebraic functions of them) which give rise to charges in
involution. In particular we note that $q_{\pm 2}$ always commutes
with the other charges, showing that all higher-spin charges are
classically in involution with energy-momentum.\vspace{1em}

We must now deal separately with the cases $G=SU(m|n)$ and
$G=OSp(m|2n)$, and we shall see that in the former case we must
impose $m\neq n$.


\subsection{Commuting Charges for $SU(m|n)$}

It is evident that $X \in su(m|n)$ does not imply $X^p \in su(m|n)$
for all integers $p$ because supertracelessness will not in general
hold. In the case $m\neq n$ we can replace $j_+^r$ by the
supertraceless and anti-hermitian
$j_+^r-(1/l)\textmd{Str}(j_+^r)\mathbb{I}_{m+n}$ where $l=m-n$, and
then using the completeness condition (\ref{completeness}) we find
\begin{equation}
\{ q_{\pm r},q_{\pm s}\}=\mp \frac{2(r+1)(s+1)\kappa}{l}\int
\textit{dx} \; \textmd{Str} (j_{\pm}^r)\partial_x \textmd{Str}
(j_{\pm}^s)
\end{equation}

This term is not in general zero. It is necessary to know the exact
form of the Poisson brackets for $\mathcal{J}_{\pm
r}=\textmd{Str}(j_{\pm}^r)$.  After some computation the result is
\begin{eqnarray}\label{su commute currents}
\{ \mathcal{J}_r(x),\mathcal{J}_s(y) \} & = & \left( \frac{rs}{
l}\mathcal{J}_{r-1}(x)\mathcal{J}_{s-1}(x)
-rs\mathcal{J}_{r+s-2}(x)\right)
\delta '(x-y)\\
& + & \left(
\frac{rs}{l}\mathcal{J}_{r-1}(x)\mathcal{J}'_{s-1}(x)-\frac{rs(s-1)}{(r+s-2)}
\mathcal{J}'_{r+s-2}(x) \right) \delta(x-y)\nonumber\end{eqnarray}

Note the similarity between this and the Poisson brackets for the
currents of the bosonic $SU(l)$ model \cite{local conserved
charges}. Indeed, these Poisson brackets are all antisymmetric, and
feature purely bosonic currents. This is also the point where we
must distinguish between the quantum conformal model $PSU(n|n)$ and
the non-conformal models.\vspace{1em}

We now want to find a set of algebraically independent currents
which give rise to a set of mutually commuting charges. We follow a
similar method to \cite{local conserved charges} and define a
generating function $A(x,\lambda)$ with a spectral parameter
$\lambda$ by
\begin{equation}\label{sdet}
A(x,\lambda)=\textmd{sdet}\left(1-\lambda
j_+(x)\right)=\exp\left(-\sum_{r=2}^{\infty}\frac{\lambda^r}{r}\mathcal{J}_r(x)
\right)
\end{equation}

and then claim that the set of currents defined by
\begin{equation}\label{su currents}
\left.\mathcal{K}_{r+1}(x)=A(x,\lambda)^{r/l}\right|_{\lambda^{r+1}}
\end{equation}

form commuting charges upon integration over space.
\begin{equation}\label{su primitive commute}
\left.\int dx \, dy \{
A(x,\mu)^{r/l},A(y,\nu)^{s/l}\}\right|_{\mu^{r+1}\nu^{s+1}} =0
\end{equation}

This differs from the bosonic model \cite{local conserved charges}
in that (\ref{sdet}) is not a polynomial of finite order in
$\lambda$ (as it is in the bosonic case), but a rational function.
For the bosonic model, when $r\equiv 0\textmd{ mod } l$ current
(\ref{su currents}) would be exactly zero. The local charges
therefore have spins equal to the exponents of the Lie algebra
modulo its Coxeter number $h$. No such pattern seems to exist for
the supergroup model, and (\ref{su currents}) seems to be non-zero
for all positive integer values of $r$.\vspace{1em}

However after noting this, we can proceed with an analogous
argument. We calculate $\{\ln A(x,\mu),\ln A(y,\nu)\}$ using
(\ref{su commute currents}), and thence it is seen that (\ref{su
primitive commute}) is satisfied.\vspace{1em}

The infinite number of currents (\ref{su currents}) are not
algebraically independent. The number of independent currents is
equal to the rank of the superalgebra, so for $su(m|n)$ there are
$m+n-1$ independent currents. Therefore a maximal set of
algebraically independent currents which form commuting charges are
\begin{equation}
\{ K_i(x) \; | \; 2\leq i\leq m+n\}.
\end{equation}

Any higher spin currents which give further commuting charges must
necessarily be algebraic functions of these currents. The form of
the currents is identical to that of $SU(m-n)$ \cite{local conserved
charges}, but we shall reproduce the first few examples of them here
for completeness.

\begin{eqnarray}
\mathcal{K}_2  &  =  &  \mathcal{J}_2  \nonumber   \\
\mathcal{K}_3  &  =  &  \mathcal{J}_3  \nonumber   \\
\mathcal{K}_4  &  =  &  \mathcal{J}_4-\frac{3}{2(m-n)}\mathcal{J}_2^2  \nonumber   \\
\mathcal{K}_5  &  =  &
\mathcal{J}_5-\frac{10}{3(m-n)}\mathcal{J}_3\mathcal{J}_2
\end{eqnarray}


\subsection{Commuting Charges for $OSp(m|2n)$}

Using the defining relation for this superalgebra, we see that $X\in
osp(m|2n)$ implies that $X^p\in osp(m|2n)$ if $p$ is odd. So for
such $p$, we can use the completeness condition (\ref{completeness})
to show that the integrand of (\ref{supertrace charge commutation})
is a total divergence, and thus that the charges commute.
\begin{equation}
\{q_{\pm r},q_{\pm s}\}=\pm \frac{2s(r+1)(s+1)\kappa}{r+s}\int dx\,
\partial_x \textmd{Str}(j_{\pm}^{r+s})=0
\end{equation}

So for the $OSp(m|2n)$ model, the set of $q_r$ defined by
(\ref{supertrace higher charge}) for $r$ odd is a set of mutually
commuting charges. (For $r$ even, these charges are exactly
zero.)\vspace{1em}

We will find that we can derive more interesting conserved charges
if we calculate the equal-time Poisson brackets for the currents
using (\ref{lightcone SPBs}).
\begin{equation}\label{osp strace currents}
\{\mathcal{J}_r(x),\mathcal{J}_s(y)\}=-rs\mathcal{J}_{r+s-2}(x)\delta'(x-y)
-\frac{rs(s-1)}{(r+s-2)}\mathcal{J}'_{r+s-2}(x)\delta(x-y)
\end{equation}

This equation holds for all $r,s\geq 1$, but is only interesting for
even $r,s$. Again we note the similarity between these relations and
those given for the bosonic orthogonal and symplectic algebras
\cite{local conserved charges}. Analogous to these models we will
find a family of commuting currents, with a free parameter $\alpha$.
We can formulate these currents through the use of generating
functions with a parameter $\lambda$. We define
\begin{equation}\label{B}
B(x,\lambda)=\textmd{sdet}(1-\sqrt{\lambda}j_+(x))=\exp\left(
-\sum_{r=1}^{\infty}\frac{\lambda^r}{2r}\mathcal{J}_{2r}(x) \right)
\end{equation}

and find that the currents defined by
\begin{equation}\label{commute osp currents}
\left.\mathcal{K}_{r+1}(x)=B(x,\lambda)^{\alpha
r}\right|_{\lambda^{(r+1)/2}}
\end{equation}

give rise to a mutually commuting set of currents
\begin{equation}\label{osp commute currents}
\left.\int dx\; dy\;\{B(x,\mu)^{\alpha r},B(y,\nu)^{\alpha s}
\}\right|_{\mu^{(r+1)/2}\nu^{(s+1)/2}}=0.
\end{equation}

The argument proceeds similarly to the unitary case. The difference
now is that we have a family of commuting charges, depending on a
free parameter $\alpha$. The resulting currents are similar in form
to the orthogonal and symplectic cases \cite{local conserved
charges}. We reproduce them here.
\begin{eqnarray}
\mathcal{K}_2  &  =  &  \mathcal{J}_2  \nonumber \\
\mathcal{K}_4  &  =  &
\mathcal{J}_4-\frac{3\alpha}{2}\mathcal{J}_2^2 \nonumber  \\
\mathcal{K}_6  &  =  &  \mathcal{J}_6-\frac{15\alpha}{4}\mathcal{J}_4\mathcal{J}_2
+\frac{25\alpha^2}{8}\mathcal{J}_2^3 \nonumber  \\
\mathcal{K}_8  &  =  &
\mathcal{J}_8-\frac{14\alpha}{3}\mathcal{J}_6\mathcal{J}_2-\frac{7\alpha}{4}\mathcal{J}_4^2
+\frac{49\alpha^2}{4}\mathcal{J}_4\mathcal{J}_2^2-\frac{343\alpha^3}{48}\mathcal{J}_2^4
\end{eqnarray}


\subsubsection{The Superpfaffian}
For $SO(2l)$ there is another conserved current which cannot be
expressed as the trace of a power of $j_+(x)$: the Pfaffian current
of spin $l$,
\begin{equation}\mathcal{P}(x)=\epsilon^{I_1 J_1 \ldots I_l J_l}(j_+)_{I_1 J_1}\ldots
(j_+)_{I_l J_l}
\end{equation}

By requiring that the charge associated to this current is in
involution with the set of commuting charges formed from traces, the
value of $\alpha$ is fixed to be $1/(2l-2)$ \cite{local conserved
charges}. A similar situation exists for $OSp(2m|2n)$, for which
there exists an analogous current, the superpfaffian current of spin
$m-n$ \cite{superpfaffian}. Its charge is in involution with the
supertrace currents fixes $\alpha$ to be $1/(2m-2n-2)$. The
superpfaffian current differs from the others already considered in
that it cannot be written in the form $d^{A_1\ldots A_p}j_{\pm
A_1}\ldots j_{\pm A_p}$, but is instead a rational
function.\vspace{1em}

Given any supermatrix written in block form
$\left(\begin{array}{cc}A&B\\C&D\end{array} \right)$ for which $D$
is invertible, we define the superpfaffian to be
\begin{equation}
\textmd{Spf}\left(
\begin{array}{cc}A&B\\C&D\end{array}\right)=
\frac{\textmd{Pfaff}(A-BD^{-1}C)}{\sqrt{\textmd{det(D)}}}=\frac{\textmd{Pfaff}(A)}
{\sqrt{\textmd{det}(D-CA^{-1}B)}}
\end{equation}

where $\textmd{Pfaff}$ is the ordinary pfaffian.\vspace{1em}

Let $j_+(x)\in osp(2m|2n)$. Every element of a Lie superalgebra is
locally conjugate to some element of the Cartan subalgebra. So there
exists $U(x)\in OSp(2m|2n)$ such that
\begin{equation}
Uj_+U^{-1}=\textmd{diag}\left( \left[\begin{array}{cc}0 &
\lambda_1\\-\lambda_1 & 0
\end{array}\right],\ldots ,\left[\begin{array}{cc}0 & \lambda_m\\-\lambda_m & 0
\end{array}\right],i\mu_1,\ldots i\mu_n,-i\mu_1,\ldots ,-i\mu_n \right)
\end{equation}
for real $\lambda_i(x)$, $\mu_j(x)$ (the weights) \cite{evans
mountain}. The currents that we are interested in can be expressed
as functions of the $m+n$ weights.\vspace{1em}

We first note that we can write (\ref{B}) as
\begin{equation}
\textmd{sdet}(1-\sqrt{\nu}j_+(x))=\frac{(1+\nu \lambda_1^2)\ldots
(1+\nu \lambda_m^2)}{(1+\nu \mu_1^2)\ldots (1+\nu \mu_n^2)}
\end{equation}

and then the supertrace currents (\ref{commute osp currents}) which
give commuting charges can be expressed in terms of the weights as
\begin{equation}
\mathcal{K}_p(x)=\left.\left[ \frac{(1+\nu \lambda_1^2)\ldots (1+\nu
\lambda_m^2)}{(1+\nu \mu_1^2)\ldots (1+\nu \mu_n^2)}
\right]^{\alpha(p-1)}\right|_{\nu^{p/2}}
\end{equation}

or equivalently as
\begin{equation}
\left.\mathcal{K}_p(x)=\exp
\left(-\alpha(p-1)\sum_{r=1}\frac{(-1)^r\nu^r}{r}\left(\sum_i
\lambda_i^{2r}-\sum_j \mu_j^{2r} \right)\right)\right|_{\nu^{p/2}}
\end{equation}

Our conserved superpfaffian current can then be written in terms of
the weights as
\begin{equation}
\mathcal{P}(x)=\textmd{spf}(j_+(x))=\left|\frac{\lambda_1\ldots
\lambda_m}{\mu_1\ldots \mu_n}\right|
\end{equation}

and our claim is that requiring $\int dx \; dy \;
\{\mathcal{K}_p(x),\mathcal{P}(y) \}=0$ for all (even) $p$ will
constrain $\alpha$ to be $(2m-2n-2)^{-1}$. We shall use the Poisson
bracket relations \begin{eqnarray} \{\lambda_i(x),\lambda_j(y)\}& =
& -4\kappa \delta_{ij}\delta'(x-y)\nonumber\\
 \{\mu_k(x),\mu_l(y)\}&=&4\kappa
\delta_{kl}\delta'(x-y)\end{eqnarray}

(where the difference in signs comes from the definition of
supertrace) to calculate (writing
$C(x)=\textmd{sdet}(1-\sqrt{\nu}j_+(x))$ and $\beta=\alpha(p-1)$ for
convenience)

$$
\int dx\,dy\,\{\mathcal{P}(x),C(y)^{\beta}\}=\int dx\,dy\,\beta
C(y)^{\beta-1}\{\mathcal{P}(x),C(y)\}$$

\begin{eqnarray}
=& 8\beta\kappa\nu\int dx\,
\left(\sum_{i=1}^m\partial_x\left(\frac{\mathcal{P}(x)}{\lambda_i(x)}\right)\frac{C(x)^{\beta}\lambda_i(x)}
{1+\nu\lambda_i^2(x)}-\sum_{k=1}^n\partial_x\left(\frac{\mathcal{P}(x)}{\mu_k(x)}\right)\frac{C(x)^{\beta}\mu_k(x)}
{1+\nu\mu_k^2(x)} \right)\nonumber\\
=&8\beta \kappa \nu \int dx\,
\left[(m-n)-\frac{\nu}{\beta}\partial_{\nu}-1-\frac{1}{2\beta}\right]\left(\partial_x
\mathcal{P}(x)\right)C(x)^{\beta}
\end{eqnarray}

We are only interested in the coefficient of $\nu^{p/2}$, and so we
replace $\nu\partial_{\nu}\mapsto p/2-1$. We then find that the term
in square brackets vanishes if and only if $\alpha=1/(2m-2n-2)$, as
required.\vspace{1em}

We immediately see that there is a similar pattern of spins for the
local charges on $OSp(m|2n)$ and $SO(m-2n)$; Saleur and Kaufmann
\cite{saleur kaufmann} have studied the similarity between the
S-matrices of the models with these symmetries.\vspace{1em}

Interestingly, for the $OSp(2n+2|2n)$ models (precisely those which
are exactly conformal) there exists no finite value of $\alpha$ for
which the superpfaffian charge commutes with all the other charges.
This does not affect their integrability properties as it is still
possible to construct a Lax pair.


\section{Non-Local Charges}

In addition to the local charges, there are two infinite sets of
conserved non-local charges, which are elements of
$(\textbf{g}\otimes \Lambda)_{\bar{0}}$ and generate a chiral
Yangian structure $Y(\textbf{g})_L \times Y(\textbf{g})_R$
\cite{local conserved charges, non-local charges}.\vspace{1em}

The full set of non-local charges are generated by the conserved
local charge
\begin{equation}
Q_A^{(0)}=\int^{\infty}_{-\infty}\textit{dx}\; j_{0A}(x)
\end{equation}

and the first non-local charge
\begin{equation}
Q_A^{(1)}=\int^{\infty}_{-\infty}\textit{dx}
j_{1A}(x)-\frac{1}{2\kappa}f^{BC}_{\;\;\;\;A}
\int^{\infty}_{-\infty}\textit{dx} \; j_{0B}(x)
\int^{x}_{-\infty}\textit{dy} \; j_{0C}(y)
\end{equation}

where conservation follows from (\ref{conserve_bianchi}).  Higher
charges are formed from commutations of these. Their construction in
\cite{non-local charges, Mackay yangian} for the bosonic charges can
be applied analogously here.\vspace{1em}

To show that all local charges are in involution with the non-local
charges, it suffices to show that the local charges commute with the
first two charges $Q_A^{(0)}$ and $Q_A^{(1)}$.\vspace{1em}

The invariance of the d-tensor can be used in a straightforward
fashion to show that
\begin{equation}
\{ q_s, Q_A^{(0)}\}=0
\end{equation}

Commutation of $Q_A^{(1)}$ is not so simple to show, but we can
proceed by using a similar argument to \cite{local conserved
charges}. We consider each of the two terms of $Q_B^{(1)}$
separately, and using the invariance property (\ref{d tensor}), we
find that the commutation of the first term gives
\begin{equation}\label{1st result}
\{q_s,\int dy \; j_{1B}(y)\}=-(s+1)\int dx \;d^{CA_1\ldots
A_s}f_{C\;\;B}^{\;\;D}j_{+A_s}\ldots j_{+A_1}j_{1D}
\end{equation}

As in \cite{local conserved charges}, when looking at the second
term we must be cautious when working with the limits of the spatial
integration. We thus integrate between $\pm L$ and then take the
limit $\L\rightarrow \infty$. We are interested in
\begin{equation}
\{q_s, \int_{-L}^L dy \int_{-L}^y dz \;
f^{CD}_{\;\;\;\;\;B}j_{0C}(y)j_{0D}(z) \}
\end{equation}

To evaluate this, we then introduce a step function,
\begin{equation}
\int_{-L}^Ldx\int_{-L}^L dy \int_{-L}^L dz \; d^{A_1\ldots
A_{s+1}}f^{CD}_{\;\;\;\;\;B}\{j_{+A_{s+1}}(x)\ldots j_{A_1}(x),
j_{0C}(y)j_{0D}(z) \}\theta(y-z)
\end{equation}

All the ultra-local terms (i.e. the $\delta (x-y)$ and $\delta
(x-z)$) vanish by invariance (\ref{d tensor}), and after some
computation (noting that the currents vanish at infinity) we are
left with
\begin{equation}\label{2nd result}
2\kappa (s+1)\int dx \; d^{CA_1\ldots A_s}f_{C\;\;B}^{\;\;D}
j_{+A_s}\ldots j_{+A_1}j_{0D}
\end{equation}

We recombine results (\ref{1st result}) and (\ref{2nd result}), and
again use the invariance property (\ref{d tensor}) to show the final
result
\begin{equation}
\{ q_s, Q_A^{(1)}\}=0
\end{equation}

The Yangian charges do not in general commute.
\begin{equation}
\{Q^{(0)}_A,Q^{(0)}_B\}=(-1)^{\chi}f_{AB}^{\;\;\;\;C}Q^{(0)}_C
\end{equation}

\begin{equation}
\{Q^{(0)}_A,Q^{(1)}_B\}=(-1)^{\chi}f_{AB}^{\;\;\;\;C}Q^{(1)}_C
\end{equation}

where $\chi=\eta_A.\eta_B+\eta_A+\eta_B$. Once equipped with the
additional structure of a (non-trivial) coproduct and counit, this
is the expected form of the super-Yangian algebra \cite{superyangian
on ads5, zhang, yangian symmetry in d=4 SYM}.\vspace{1em}


\section{Remarks on the Quantum Model}

To determine whether the higher spin local currents are also
conserved in the quantum theory, we use the anomaly counting method
of Goldschmidt and Witten \cite{Goldschmidt Witten}. This is a
rather indirect method, and does not convey an insight into the form
of these anomalies. Instead it tries to show that any quantum
anomalies can be written in the form of a total derivative, in which
case a conservation law still exists, although in a modified
form.\vspace{1em}

We must consider all possible anomaly terms which have the same
behaviour under the symmetries of the model. These symmetries
comprise of the continuous Lorentz and chiral symmetries, and also
some discrete symmetries.\vspace{1em}

For any supergroup, the principal chiral model is invariant under
the map $\pi: g\mapsto g^{-1}$. This exchanges the left and right
currents. Additional symmetries arise as outer autmorphisms of the
underlying Lie superalgebras \cite{sorba}.
\begin{eqnarray}
\gamma: g\mapsto g^* & \textmd{for }\mathbf{g}=su(m|n)\\
\sigma: g\mapsto MGM^{-1} & \textmd{for }\mathbf{g}=osp(2m|2n)
\end{eqnarray}

where $M$ is an element of $OSP(2m|2n)$ with superdeterminant -1.
Each of the currents $\mathcal{J}_r(x)$ are either odd or even under
the action of all of these symmetries. The argument proceeds
identically to the bosonic PCM \cite{local conserved charges}, and
we shall not reproduce it all here, but instead illustrate the idea
with the spin 2 example. It does not matter for the following
whether we are considering $su(m|n)$ or $osp(m|2n)$; the results are
the same for all models.\vspace{1em}

There are only two spin 2 currents,
$\mathcal{J}_2=\textmd{Str}(j_{\pm}^2)$. (We shall just consider the
+ current, the argument is identical for the other.) This is even
under the discrete symmetries, invariant under the chiral symmetry,
and of mass dimension 2. There is only one possible anomaly term
with identical behaviour, $\textmd{Str}(j_{-}\partial_{+}j_{+})$,
which can be written as a total derivative
$\partial_{+}\textmd{Str}(j_-j_+)$. So the only possible correction
to the conservation law is
\begin{equation}
\partial_-\mathcal{J}_2=\alpha \partial_{+}\textmd{Str}(j_-j_+)
\end{equation}

where $\alpha$ is some unknown parameter, and we have a modified
conservation law. (Note that $\alpha$ can be zero, and we would
expect it to be so for those models which retain quantum conformal
invariance.)\vspace{1em}

Similar results hold for spin 3 and 4 currents, but not for higher
spins. This does not mean that there is no quantum conserved
current, for the anomaly counting method is sufficient but not
necessary. Nevertheless, integrability is guaranteed by the
existence of one higher spin conservation law \cite{parke}.


\section{Conclusions and Further Questions}

We have derived a set of commuting charges for the principal chiral
model on the Lie supergroups $SU(m|n)$ for $m\neq n$, and
$OSp(m|2n)$. These are integrals of local currents, each constructed
with the use of an invariant of the underlying Lie
superalgebra.\vspace{1em}

The $SU(m|n)$ models have conserved currents generated by $m+n-1$
primitive currents. The current algebra is similar in appearance to
that of $SU(m-n)$.\vspace{1em}

The orthosymplectic models $OSp(2m|2n)$ and $OSp(2m+1|2n)$ have each
$m+n$ primitive currents from which the infinite charges can be
calculated. For the former set of models, the superpfaffian charge
will only commute with the other charges if $m\neq n+1$. The algebra
of currents suggests a relationship between $OSp(m|2n)$ and
$SO(m-2n)$ models \cite{saleur kaufmann}.\vspace{1em}

In the bosonic model there is a correlation between the degrees of
the primitive currents and the exponents of the underlying Lie
algebra \cite{local conserved charges}. This same pattern of
currents exists for the affine Toda field theories \cite{wilson,
olive, corrigan, 0101231}. It is therefore natural to wonder whether
a similar pattern exists between the local charges of the PCM on a
supergroup and those of the affine Toda field theory on a Lie
superalgebra \cite{evans madsen, evans hollowood}. The difficulty
here is that to each Lie superalgebra there may be associated more
than one affine Toda field theory, depending on the choice of
inequivalent simple root system. It would be interesting to consider
whether the PCM shares any properties with these models.\vspace{1em}

Quantum conformal invariance is not expected for general Lie
supergroups, but it should be possible to introduce a WZ term which
guarantees this at a certain critical limit. The integrability of
these models through a consideration of their local and non-local
conserved charges will be the scope of future work.\\

Similar to the arguments in this paper, the sigma model with a
supercoset target (either with or without a WZ term) should exhibit
local conserved charges. The construction of a set of (classically)
commuting charges for these models has yet to be
explored.\vspace{1em}

For the PCM on a Lie Group in the presence of a boundary (i.e. on the
half-line $x<0$) a natural connection arises between boundary
integrability and symmetric spaces \cite{boundary}. Future work will
investigate the analogous boundary conditions to ensure integrability on
a supergroup.\vspace{1em}

Finally the particle spectrum of these models remains to be explored.
Because of the additional complications of the
representation theory (i.e. atypicality \cite{sorba}),
this is expected to be more complicated for supergroups than for the
bosonic models.\vspace{1em}

\textbf{Acknowledgements}\\

I would like to thank Niall MacKay for his advice and supervision,
and Charles Young for useful discussions. I also thank EPSRC and the
EUCLID network, ref. HPRN-CT-2002-00325, for funding.



\begin{thebibliography}{99}

\bibitem{ads2} N. Berkovits, M. Bershadsky, T. Hauer, S. Zhukov, B.
Zwiebach, \textit{Superstring Theory on $AdS_2\times S^2$ as a Coset
Supermanifold}, Nucl. Phys. \textbf{B567} (2000) 61 [hep-th/9907200]

\bibitem{ads5} I. Bena, J. Polchinski, R. Roiban, \textit{Hidden
Symmetries of the $AdS_5\times S^5$ Superstring}, Phys. Rev.
\textbf{D69} (2004) 046002 [hep-th/0305116]

\bibitem{transition matrices} A. Das, J. Maharana, A. Melikyan, M. Sato,
\textit{The Algebra of Transition Matrices for the $AdS_5\times S^5$
Superstring}, JHEP {\bf 0412} (2004) 055 [hep-th/0411200]

\bibitem{metsaev} R. R. Metsaev, A. A. Tseytlin, \textit{Type IIB Superstring Action
in $AdS_5\times S^5$ Background}, Nucl.\ Phys.\ B {\bf 533} (1998)
109 [hep-th/9805028]

\bibitem{Zm grading} C.A.S. Young, \textit{Non-local charges,
$\mathbb{Z}_m$ Gradings and Coset Space Actions}, Phys.\ Lett.\ B
{\bf 632} (2006) 559 [hep-th/0503008]

\bibitem{Young Kagan} D. Kagan, C.A.S. Young, \textit{Conformal Sigma-Models
on Supercoset Targets}, [hep-th/0512250]

\bibitem{CFT of AdS with RR flux} N. Berkovits, C. Vafa, E. Witten,
\textit{Conformal Field Theory of AdS Background with Ramond-Ramond
Flux}, JHEP {\bf 9903} (1999) 018 [hep-th/9902098]

\bibitem{psl cft} M. Bershadsky, S. Zhukov and A. Vaintrob,
\textit{$PSl(n|n)$ sigma model as a conformal field theory}, Nucl.\
Phys.\ B {\bf 559} (1999) 205 [hep-th/9902180]

\bibitem{local conserved charges} J.M. Evans, M. Hassan, N.J. MacKay and A.J. Mountain ,
\textit{Local conserved charges in principal chiral models}, Nucl.\
Phys.\ B {\bf 561} (1999) 385 [hep-th/9902008]; \textit{Conserved
Charges and Supersymmetry in Principal Chiral and WZW models},
Nucl.\ Phys.\ B {\bf 580} (2000) 605 [hep-th/0001222]

\bibitem{evans mountain} J.M. Evans, A.J. Mountain, \textit{Commuting
Charges and Symmetric Spaces}, Phys.\ Lett.\ B {\bf 483} (2000) 290
[hep-th/0003264]

\bibitem{quantum symmetric models} J.M. Evans, D. Kagan, N.J. MacKay,
C.A.S. Young, \textit{Quantum, Higher-spin, Local Charges in
Symmetric Space Sigma Models}, JHEP {\bf 0501} (2005) 020
[hep-th/0408244]

\bibitem{saleur kaufmann} H. Saleur, B. Kaufmann, \textit{Integrable Quantum Field Theories
with $OSp(m|2n)$ Symmetries}, Nucl. Phys. \textbf{B628} (2002) 407
[hep-th/0112095]

\bibitem{polchinkski supercoset} N. Mann, J. Polchinski, \textit{Bethe Ansatz for
a Quantum Supercoset Sigma Model}, Phys.\ Rev.\ D {\bf 72} (2005)
086002 [hep-th/0508232]

\bibitem{superpfaffian} P. Lavaud, \textit{Superpfaffian},
[math.GR/0402067]

\bibitem{Goldschmidt Witten} Y.Y. Goldschmidt, E. Witten, \textit{Conservation Laws in Some
Two-Dimensional Models}, Phys. Lett. \textbf{B91} (1980) 392

\bibitem{superyangian on ads5}  M. Hatsuda and K. Yoshida, \textit{Classical Integrability
and Super Yangian on $AdS_5 \times S^5$}, [hep-th/0407044]

\bibitem{zhang} R. Zhang, \textit{Representations of Super Yangians}, J. Math. Phys. 36
(1995) 3854; \textit{The $gl(M|N)$ Super Yangian and its Finite
Dimensional Representations}, Lett. Math. Phys. \textbf{37} (1996)
419

\bibitem{yangian symmetry in d=4 SYM} L. Dolan, C. Nappi, E. Witten,
\textit{Yangian Symmetry in D=4 Superconformal Yang-Mills Theory},
[hep-th/0401243]

\bibitem{sorba} A. Frappat, A. Sciarrino and P. Sorba, \textit{Structure of Basic Lie
Superalgebras and of their Affine Extensions}, Commun. Math. Phys.
\textbf{121} (1989) 457; \textit{Dictionary on Lie Algebras and
Superalgebras}, Academic Press, 2000

\bibitem{Kac} V. G. Kac, \textit{Lie Superalgebras}, Adv. in Maths.
\textbf{26} (1977) 8

\bibitem{scheunert} Scheunert, \textit{Lie Superalgebras},
 Lecture Notes in Mathematics 716, Springer-Verlag, 1979

\bibitem{superpoisson} J. A. Azc\'arraga, J. M. Izquierdo, A. M.
Perelomov and J. C. P\'rez Bueno, \textit{The $Z_2$-graded
Schouten-Nijenhuis bracket and generalised super-Poisson
structures}, J.\ Math.\ Phys.\  {\bf 38} (1997) 3735
[hep-th/9612186]

\bibitem{faddeev takhtajan} L.D. Faddeev, L.A. Takhtajan, \textit{Hamiltonian Methods in the Theory of
Solitons}, Springer Series in Soviet Mathematics, 1987

\bibitem{invariant simple} J. A. de Azc\'arraga, A. J. Macfarlane,
A. J. Mountain and J. C. P\'erez Bueno, \textit{Invariant tensors
for simple groups}, Nucl.\ Phys.\ B {\bf 510} (1998) 657
[physics/9706006]

\bibitem{invariant compact} A. J. Mountain, \textit{Invariant tensors and Casimir operators
for simple compact Lie groups}, J.\ Math.\ Phys.\ {\bf 39} (1998)
5601

\bibitem{Casimir invariants and characteristic identities} P. D.
Jarvis, H. S. Green, \textit{Casimir invariants and characteristic
identities for generators of the general linear, special linear and
orthosymplectic graded Lie algebras}, J. Math. Phys \textbf{20}
(1979) 2115

\bibitem{invariant polynomials} A. Sergeev, \textit{The invariant polynomials on simple
Lie superalgebras}, math.RT/9810111, \textit{An analog of the
classical invariant theory for Lie superalgebras}, [math.RT/9810113]

\bibitem{non-local charges} E. Br\'ezin, C. Itzykson, J.
Zinn-Justin and J.B. Zuber, \textit{Remarks on the existence of
non-local charges in two-dimensional methods}, Phys. Lett.
\textbf{82B} (1979) 442

\bibitem{Mackay yangian} N. J. MacKay, \textit{Introduction to Yangian symmetry in
integrable field theory}, Int.\ J.\ Mod.\ Phys.\ A {\bf 20} (2005)
7189 [hep-th/0409183]

\bibitem{parke} S. Parke, \textit{Absence of particle production and factorization of the
S-matrix in 1+1-D models}, Nucl. Phys. \textbf{B174} (1980) 166

\bibitem{wilson} G. Wilson, \textit{The Modified Lax and Two-Dimensional Toda Lattice Equations
Associated with Simple Lie Algebras}, Ergod. Th. Dynam. Sys.
\textbf{1} (1981) 361

\bibitem{olive} D. Olive, N. Turok, \textit{Local Conserved Densities and Zero-Curvature Conditions
for Toda Lattice Field Theories}, Nucl. Phys. \textbf{B257} (1985)
277

\bibitem{corrigan} E. Corrigan, \textit{Recent Developments in Affine Toda Quantum Field
Theory}, [hep-th/9412213]

\bibitem{0101231}J.M. Evans, \textit{Integrable sigma-models and
Drinfeld-Sokolov Hierarchies}, Nucl.\ Phys.\ B {\bf 608} (2001) 591
[hep-th/0101231]

\bibitem{evans madsen} J. Evans, J.O. Madsen, \textit{Dynkin Diagrams and Integrable Models
based on Lie Superalgebras}, Nucl.\ Phys.\ {\bf B503} (1997) 715
[hep-th/9703065]

\bibitem{evans hollowood} J. Evans, T. Hollowood, \textit{Supersymmetric Toda field theories},
Nucl.\ Phys.\ {\bf B352} (1991) 723

\bibitem{boundary} N.J. MacKay, B.J. Short, \textit{Boundary Scattering,
Symmetric Spaces and the Principal Chiral Model on the Half-line},
Commun.\ Math.\ Phys.\  {\bf 233} (2003) 313 [hep-th/0104212]




\end{thebibliography}
\end{document}